\documentclass[a4paper, 11pt, twocolumn]{article}
\usepackage[a4paper,margin=2cm]{geometry}

\usepackage[
backend=biber,
sorting=none,
citestyle=phys,
style=nature
]{biblatex}
\usepackage{tikz}
\usetikzlibrary{arrows}
\usepackage{booktabs}
\usetikzlibrary{calc}
\usepackage{amsmath}
\usepackage[cleanup]{gnuplottex}
\usepackage{wasysym}
\usepackage{xparse}
\usepackage{lineno}
\ExplSyntaxOn
\DeclareExpandableDocumentCommand{\convertlen}{ O{cm} m }
 {
  \dim_to_decimal_in_unit:nn { #2 } { 1 #1 } cm
 }
\ExplSyntaxOff

\DeclareMathOperator\erf{erf}

\usepackage{graphicx}

\makeatletter
\newcommand{\p}{\vec{p}\@ifnextchar{^}{\,}{}}
\renewcommand{\d}{\text{d}}
\makeatother

\title{Timing detectors for forward physics}
\author{Rafał Staszewski and Janusz J. Chwastowski\\[2ex]
Institute of Nuclear Physics 
Polish Academy of Sciences\\
Radzikowskiego 152, 
31--342 Kraków, Poland
}

\date{\today}

\renewcommand{\L}{\text{L}}
\newcommand{\R}{\text{R}}

\begin{document}
\twocolumn[
\begin{@twocolumnfalse}
  \maketitle
  \hrule
  \begin{abstract}
    The use of precise time-of-flight (ToF) detectors for measurements of diffractive and electromagnetic processes in proton--proton collisions is discussed.
    The performance of background rejection exploiting the ToF measurements of the forward protons is derived.
    Influence of additional timing measurements delivered by the central detectors is presented and the possible gain is studied.
  \end{abstract}
  \vspace{1ex}

  \hrule
  \vspace{2em}
\end{@twocolumnfalse}
]

\section{Introduction}
Measurements of diffractive and electromagnetic processes by means of detection of forward protons (or anti-protons) have been performed at various high-energy hadron accelerators.
They are also a part of the physics programme pursued at the Large Hadron Collider, where two beam intersection regions have been equipped with detectors allowing measurements of protons scattered at very small angles \cite{alfa, afp, ctpps, totem}.

The area of interest covers a broad range of processes starting from elastic scattering and soft diffraction, through hard diffraction, up to central exclusive production and photon--photon processes.
Measurements of processes with high cross sections are typically performed in special runs with low instantaneous luminosity, often with a dedicated accelerator optics.
However, when low cross sections are of interest, the measurement has to be performed in standard LHC conditions.

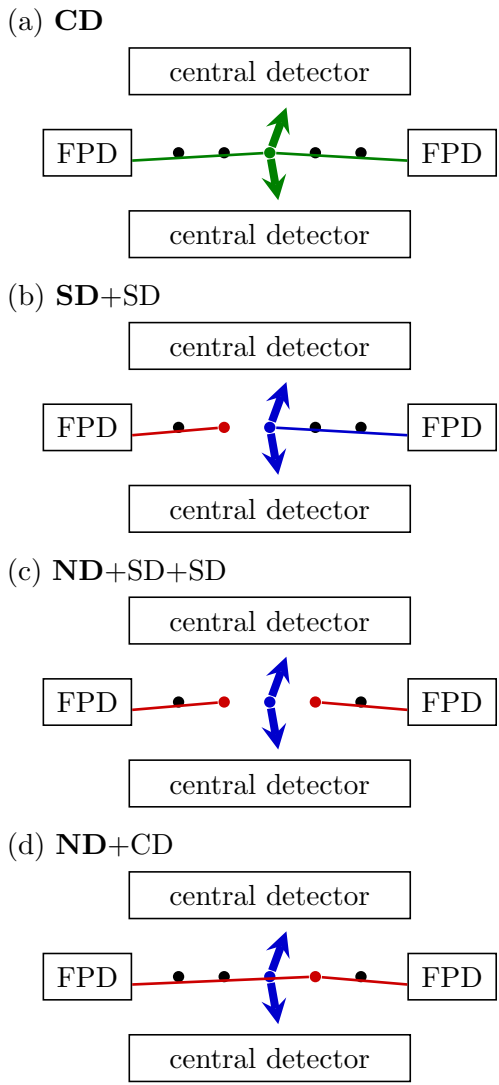
\begin{figure}[htb]
  \centering

  \newcommand{\FPDx}{6}
  \newcommand{\FPDy}{0}
  \newcommand{\CDt}{2.7}
  \newcommand{\CDb}{2.7}
  \newcommand{\PS}{1.5pt}
  \newcommand{\scol}{green!50!black}
  \newcommand{\bcola}{red!80!black}
  \newcommand{\bcolb}{blue!80!black}
  \newcommand{\scale}{0.4}

  \begin{tikzpicture}[style=semithick, scale=\scale]
    \node[anchor=west] at (-9,4.3) {(a) \textbf{CD}};
    \node[circle,fill,inner sep=\PS] at (-1.5,0) (A) {};
    \node[circle,fill=\scol,inner sep=\PS] at ( 0,0) (B) {};
    \node[circle,fill,inner sep=\PS] at ( 1.5,0) (C) {};
    \node[circle,fill,inner sep=\PS] at ( 3,0) {};
    \node[circle,fill,inner sep=\PS] at (-3,0) {};
    \node[draw, inner sep=5pt] at (-\FPDx, \FPDy) (L) {FPD};
    \node[draw, inner sep=5pt] at (+\FPDx, \FPDy) (R) {FPD};
    \node[draw, inner sep=5pt, inner xsep=15pt] at (0, +\CDt) (T) {central detector};
    \node[draw, inner sep=5pt, inner xsep=15pt] at (0, -\CDb) (T) {central detector};
    \draw[->,>=stealth, \scol, line width=3pt] (B) -- +(70: 1.6);
    \draw[->,>=stealth, \scol, line width=3pt] (B) -- +(-80: 1.6);
    \draw[\scol, line width=1pt] (L.-10) -- (B);
    \draw[\scol, line width=1pt] (R.-170) -- (B);
  \end{tikzpicture}
  \\[1ex]
  \begin{tikzpicture}[style=semithick, scale=\scale]
    \node[anchor=west] at (-9,4.3) {(b) \textbf{SD}+SD};
    \node[circle,fill=\bcola,inner sep=\PS] at (-1.5,0) (A) {};
    \node[circle,fill=\bcolb,inner sep=\PS] at ( 0,0) (B) {};
    \node[circle,fill,inner sep=\PS] at ( 1.5,0) (C) {};
    \node[circle,fill,inner sep=\PS] at ( 3,0) {};
    \node[circle,fill,inner sep=\PS] at (-3,0) {};
    \node[draw, inner sep=5pt] at (-\FPDx, \FPDy) (L) {FPD};
    \node[draw, inner sep=5pt] at (+\FPDx, \FPDy) (R) {FPD};
    \node[draw, inner sep=5pt, inner xsep=15pt] at (0, +\CDt) (T) {central detector};
    \node[draw, inner sep=5pt, inner xsep=15pt] at (0, -\CDb) (T) {central detector};
    \draw[->,>=stealth, \bcolb, line width=3pt] (B) -- +(70: 1.6);
    \draw[->,>=stealth, \bcolb, line width=3pt] (B) -- +(-80: 1.6);
    \draw[\bcola, line width=1pt] (L.-10) -- (A);
    \draw[\bcolb, line width=1pt] (R.-170) -- (B);
  \end{tikzpicture}
  \\[1ex]
  \begin{tikzpicture}[style=semithick, scale=\scale]
    \node[anchor=west] at (-9,4.3) {(c) \textbf{ND}+SD+SD};
    \node[circle,fill=\bcola,inner sep=\PS] at (-1.5,0) (A) {};
    \node[circle,fill=\bcolb,inner sep=\PS] at ( 0,0) (B) {};
    \node[circle,fill=\bcola,inner sep=\PS] at ( 1.5,0) (C) {};
    \node[circle,fill,inner sep=\PS] at ( 3,0) {};
    \node[circle,fill,inner sep=\PS] at (-3,0) {};
    \node[draw, inner sep=5pt] at (-\FPDx, \FPDy) (L) {FPD};
    \node[draw, inner sep=5pt] at (+\FPDx, \FPDy) (R) {FPD};
    \node[draw, inner sep=5pt, inner xsep=15pt] at (0, +\CDt) (T) {central detector};
    \node[draw, inner sep=5pt, inner xsep=15pt] at (0, -\CDb) (T) {central detector};
    \draw[->,>=stealth, \bcolb, line width=3pt] (B) -- +(70: 1.6);
    \draw[->,>=stealth, \bcolb, line width=3pt] (B) -- +(-80: 1.6);
    \draw[\bcola, line width=1pt] (L.-10) -- (A);
    \draw[\bcola, line width=1pt] (R.-170) -- (C);
  \end{tikzpicture}
  \\[1ex]
  \begin{tikzpicture}[style=semithick, scale=\scale]
    \node[anchor=west] at (-9,4.3) {(d) \textbf{ND}+CD};
    \node[circle,fill,inner sep=\PS] at (-1.5,0) (A) {};
    \node[circle,fill=\bcolb,inner sep=\PS] at ( 0,0) (B) {};
    \node[circle,fill=\bcola,inner sep=\PS] at ( 1.5,0) (C) {};
    \node[circle,fill,inner sep=\PS] at ( 3,0) {};
    \node[circle,fill,inner sep=\PS] at (-3,0) {};
    \node[draw, inner sep=5pt] at (-\FPDx, \FPDy) (L) {FPD};
    \node[draw, inner sep=5pt] at (+\FPDx, \FPDy) (R) {FPD};
    \node[draw, inner sep=5pt, inner xsep=15pt] at (0, +\CDt) (T) {central detector};
    \node[draw, inner sep=5pt, inner xsep=15pt] at (0, -\CDb) (T) {central detector};
    \draw[->,>=stealth, \bcolb, line width=3pt] (B) -- +(70: 1.6);
    \draw[->,>=stealth, \bcolb, line width=3pt] (B) -- +(-80: 1.6);
    \draw[\bcola, line width=1pt] (L.-10) -- (C);
    \draw[\bcola, line width=1pt] (R.-170) -- (C);
  \end{tikzpicture}

  \caption{Jet events with two forward protons: (a) central diffractive jets, (b) single diffractive jets + single diffraction, (c) non-diffractive jets + two single diffraction processes, (d) non-diffractive jets + central diffraction.}
  \label{fig:CDevents}
\end{figure}

High instantaneous luminosity of LHC leads to high pile-up, \textit{i.e.} many independent proton--proton interactions occurring in the same experimentally registered event (bunch crossing).
The presence of pile-up poses a problem for measurements of processes with forward protons by introducing a combinatorial background.
Let us consider an event containing a pair of jets registered in the central detectors (trackers and calorimeters) and two forward protons registered in forward proton detectors (FPD). 
Such a configuration characterises a central diffractive event (\textbf{CD}),  see Fig. \ref{fig:CDevents}(a). 
However, since the forward protons can be produced with large cross sections in soft processes, mainly single diffractive dissociation and central diffraction, the same experimental signature -- two jets plus two protons -- can be obtained also from other processes as an overlay of:
\begin{itemize}
  \item single diffractive production of jets (producing two jets and one proton) plus a proton from another pile-up interaction 
    (\textbf{SD}+SD), see Fig. \ref{fig:CDevents}(b),
  \item non-diffractive production of jets (producing two jets) plus two protons from two pile-up interactions (\textbf{ND}+SD+SD), see Fig. \ref{fig:CDevents}(c),
  \item non-diffractive production of jets plus two protons from another pile-up interaction (\textbf{ND}+CD), see Fig. \ref{fig:CDevents}(d).
\end{itemize}

The non-diffractive production of jets is characterised by a much higher cross section than that of the single diffractive jet production, which in turn has higher cross section than the central diffractive jet production.
Therefore, with increasing number of $pp$ interactions in a registered  event, it becomes much more probable (even by orders of magnitude, see \cite{ATL-PHYS-PUB-2015-003}) to observe a background than a signal event.
It has to be stressed that the jet production processes are used here merely as an example.
Analogous backgrounds are present to processes other than diffractive production of jets.
The results presented in this paper are universal, that is they are equally applicable to other channels.

One of the most important ways to reject the backgrounds to central diffractive processes makes use of precise timing detectors.
This method compares the measured arrival times of the forward protons and the measured longitudinal position of the  central interaction (vertex).
The goal of this paper is to discuss a possible gain in the background discrimination due to a precise measurement of the arrival time of forward protons.
In addition, it is shown how such measurements can be combined with timing measurements performed on the central state and a possible gain is estimated.

\section{Timing measurement of forward protons}

Several feasibility studies assuming  the timing measurements of the forward protons have already been performed for different channels, see for example \cite{Cox:2007sw,Albrow:2008pn} or more recent \cite{Harland-Lang:2018hmi}.
These studies were performed assuming certain experimental conditions.
In the present analysis, the aim is to consider a general case and find how the background rejection depends on the assumed conditions.

In a classic proton--proton collider experiment, the beams are structured and consist of bunches of particles.
An experimental event occurs when a bunch from one beam passes through a bunch belonging to the other beam.
Assuming a longitudinal Gaussian structure of a bunch with the width $\sigma_b$
\renewcommand{\b}{\text{b}}
\begin{linenomath*}\[
  \rho_\b(z) \sim e^{-z^2/(2\sigma^2_\b)},
  \]\end{linenomath*}
the distributions of time and position of the $pp$ interaction vertex  can be easily calculated by multiplying densities of two 
bunches moving with approximately the speed of light (using natural units with $c=1$):
\begin{multline*}
  \rho(z, t) 
  \sim 
  e^{-(z-t)^2/(2\sigma^2_\b)}
  \cdot
  e^{-(z+t)^2/(2\sigma^2_\b)} \\
  = 
  e^{-z^2/\sigma^2_\b}
  \cdot
  e^{-t^2/\sigma^2_\b}.
\end{multline*}
One can see that both $t$ and $z$ variables follow Gaussian distributions with widths of $\sigma_\b/\sqrt{2}$.
It is worth noting that in the case of non-Gaussian bunch distributions, the $t$ and $z$ distributions do not have to be completely independent.

Since $pp$ interactions are distributed both in $t$ and in $z$, the time measurement of a single forward proton is not sufficient to constrain the position of the interaction vertex.
Therefore, the standard approach is to combine the measurements of the arrival times of both forward protons.
For a signal (central diffractive) event, assuming that the interaction took place at $(t_0, z_0)$, the measured proton arrival time on the side with positive $z$ is:
\begin{linenomath*}\[
  t_+ =  t_{0} - z_{0} + L_{+} + \delta_+,
  \]\end{linenomath*}
where $L_{+}$ is the position of the forward proton detector from $z=0$ and $\delta_+$ is the  measurement error.
For the negative-$z$ side:
\begin{linenomath*}\[
  t_- =  t_{0} + z_{0} + L_{-} + \delta_-.
  \]\end{linenomath*}
The difference of arrival times of the two protons
\begin{linenomath*}\[
  \Delta t = t_- - t_+ = 2\cdot z_0 - L_{+} + L_{-} - \delta_+ + \delta_-
  \]\end{linenomath*}
does not depend on $t_0$ anymore and, after correcting for the positions of the detectors, it can be compared to the interaction vertex position measured directly by the central detectors:
\begin{linenomath*}\[
  \Delta z = z_\text{vertex} - z_\text{time} = \delta_z + (\delta_+ - \delta_-)/2.
  \]\end{linenomath*}
$\Delta z$ is equal to zero up to the measurement errors of the arrival times ($\delta_+$, $\delta_-$) and the vertex position ($\delta_z$).
Assuming equal resolutions of both timing detectors:
\begin{linenomath*}\[
  \sigma^2(\delta_+) = \sigma^2(\delta_-) =
  \sigma^2_p
  \]\end{linenomath*}
the $\Delta z$ distribution variance is given by:
\begin{linenomath*}\[
  \sigma^2(\Delta z) = \sigma^2_p/2 + \sigma^2_z,
  \]\end{linenomath*}
where $\sigma^2_z$ is the resolution of the vertex position measurement.

In a similar way, one can define the times measured by both detectors in all four considered event types and calculate the widths of 
the resulting $\Delta z$ distributions. Results of these calculations are presented in Table \ref{tab:times}, 
where index $0$ corresponds to the $pp$ interaction from which the central state (\textit{e.g.} jets) originates
and indices $1$ and $2$ correspond to pile-up interactions in which the forward protons are produced.

\begin{table*}[hbtp]

  \setlength{\arraycolsep}{3ex}
  \renewcommand{\arraystretch}{1.4}
  \caption{Values of arrival times measured by forward proton detectors for signal and background events and the resulting variances of $\Delta z$ distributions.}
  \label{tab:times}

  \begin{linenomath*}\[
    \begin{array}[]{l  l  l  l  l |}
      \toprule
      \text{Process} &
      \text{Measured time}\ t_+ & 
      \text{Measured time}\ t_- & 
      \text{Variance}\ \sigma^2(\Delta z)
      \\ 
      \midrule
      \text{\textbf{CD}} &
      L_{+} + \delta_+ + t_{0} - z_{0} &
      L_{-} + \delta_- + t_{0} + z_{0} &
      \sigma^2_p/2 + \sigma^2_z
      \\
      \text{\textbf{SD}+SD} &
      L_{+} + \delta_+ + t_{0} - z_{0} &
      L_{-} + \delta_- + t_{1} + z_{1} &
      \sigma^2_p/2 + \sigma^2_z + \sigma^2_\b/2

      \\
      \text{\textbf{ND}+SD+SD} &
      L_{+} + \delta_+ + t_{1} - z_{1} &
      L_{-} + \delta_- + t_{2} + z_{2} &
      \sigma^2_p/2 + \sigma^2_z + \sigma^2_\b
      \\
      \text{\textbf{ND}+CD} &
      L_{+} + \delta_+ + t_{1} - z_{1} &
      L_{-} + \delta_- + t_{1} + z_{1} &
      \sigma^2_p/2 + \sigma^2_z + \sigma^2_\b
      \\
      \bottomrule
    \end{array}
    \]\end{linenomath*}

\end{table*}

The formulae for variances reveal why $\Delta z$ is a useful variable for the signal--background event discrimination. 
One can immediately see that for the signal process, the width of the distribution is given only by resolutions of the arrival time 
measurements and the vertex reconstruction resolution.
For all backgrounds, the widths are increased by the corresponding spreads of the time and position of the interactions.

\begin{figure}[htbp]
  \centering
  \includegraphics[width=\linewidth,page=1]{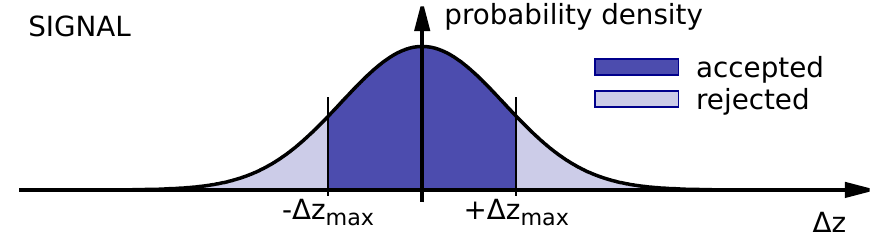}\\
  \includegraphics[width=\linewidth,page=2]{sketch.pdf}
  \caption{Illustration of signal and background selection. Darker areas correspond to the $|\Delta z| < \Delta z_\text{max}$ condition.}
  \label{fig:sketch}
\end{figure}

Qualitatively speaking, a good background rejection is possible if the experimental resolutions ($\sigma_p$ and $\sigma_z$) are significantly smaller than the bunch width $\sigma_b$.
Quantitatively, one always deals with an interplay between the fraction ($f$) of the accepted signal and background events, see Figure \ref{fig:sketch}.
Assuming normally distributed variables, the fraction of events with $|\Delta z| < \Delta z_\text{max}$ can is given by :
\begin{linenomath*}
  \begin{multline*}
    f\left(\Delta z_\text{max}\right) = \int_{-\Delta z_\text{max}}^{+\Delta z_\text{max}} \frac{1}{\sqrt{2\pi}\sigma} e^{-\Delta z^2/(2\sigma^2)} 
    \,\text{d}\Delta z \\
    =
    \erf\left(\frac{\Delta z_\text{max}}{\sqrt{2}\sigma}\right).
  \end{multline*}
\end{linenomath*}
Selecting signal and background events using the same value of $\Delta z_\text{max}$ results in:
\begin{linenomath*}\[
  f_\text{accepted bkg.} = \erf\left( \frac{\sigma_S}{\sigma_B} \erf^{-1} f_\text{accepted sig.} \right)
  \]\end{linenomath*}
where $\sigma_S$ and $\sigma_B$ are the distribution widths of signal and background, respectively.

\begin{figure}[htbp]
  \centering
  \includegraphics[width=\linewidth,page=1]{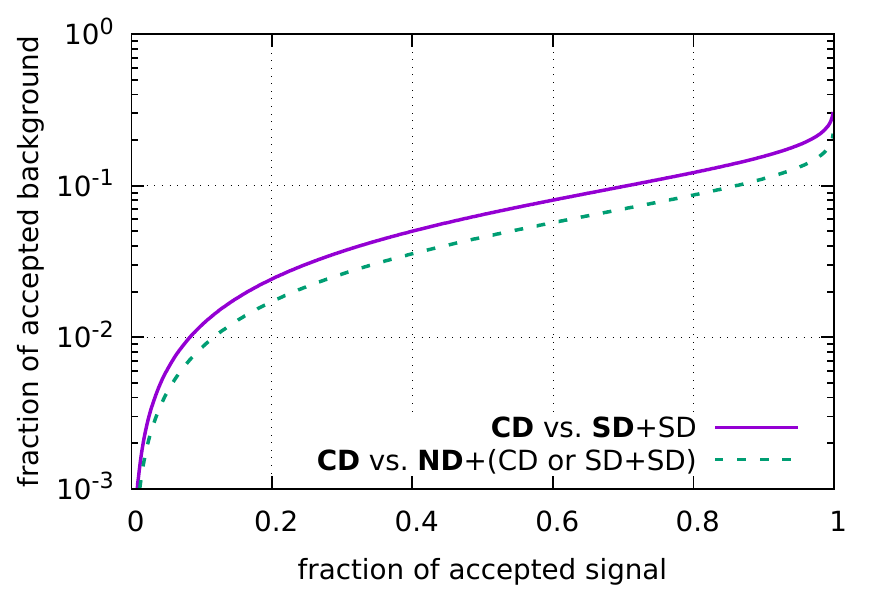}
  \caption{Performance of the signal vs. background discrimination using $\Delta z$ variable.}
  \label{fig:ROCdz}
\end{figure}

Figure \ref{fig:ROCdz} presents an example of the above dependence for the \textbf{CD} signal and the discussed backgrounds 
using the LHC-inspired assumptions:
\begin{itemize}
  \item the LHC bunch width of 75.5 mm \cite{LHCDR}, 
  \item the reconstruction resolution of the interaction vertex $z$-coordinate of 1 mm \cite{ATL-PHYS-PUB-2015-026}, 
  \item a 30 ps timing resolution of FPDs \cite{afp}.
\end{itemize}
For a reasonable choice of 90\% of accepted signal, the ND background is reduced by a factor of about 9.
Allowing the loss of additional 10\% of signal increases the rejection to a factor of about 12.
For the SD process, the rejection is slightly worse -- a factor of about 6 and about 8, for signal efficiency of 90\% and 80\%, respectively.

\section{Timing measurement in central detectors}

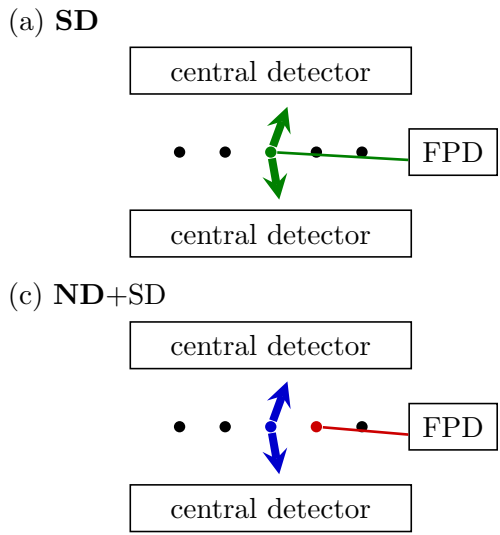
\begin{figure}[htb]
  \centering

  \newcommand{\FPDx}{6}
  \newcommand{\FPDy}{0}
  \newcommand{\CDt}{2.7}
  \newcommand{\CDb}{2.7}
  \newcommand{\PS}{1.5pt}
  \newcommand{\scol}{green!50!black}
  \newcommand{\bcola}{red!80!black}
  \newcommand{\bcolb}{blue!80!black}
  \newcommand{\scale}{0.4}

  \begin{tikzpicture}[style=semithick, scale=\scale]
    \node[anchor=west] at (-9,4.3) {(a) \textbf{SD}};
    \node[circle,fill,inner sep=\PS] at (-1.5,0) (A) {};
    \node[circle,fill=\scol,inner sep=\PS] at ( 0,0) (B) {};
    \node[circle,fill,inner sep=\PS] at ( 1.5,0) (C) {};
    \node[circle,fill,inner sep=\PS] at ( 3,0) {};
    \node[circle,fill,inner sep=\PS] at (-3,0) {};
    \node[draw, inner sep=5pt] at (+\FPDx, \FPDy) (R) {FPD};
    \node[draw, inner sep=5pt, inner xsep=15pt] at (0, +\CDt) (T) {central detector};
    \node[draw, inner sep=5pt, inner xsep=15pt] at (0, -\CDb) (T) {central detector};
    \draw[->,>=stealth, \scol, line width=3pt] (B) -- +(70: 1.6);
    \draw[->,>=stealth, \scol, line width=3pt] (B) -- +(-80: 1.6);
    \draw[\scol, line width=1pt] (R.-170) -- (B);
  \end{tikzpicture}
  \\[1ex]
  \begin{tikzpicture}[style=semithick, scale=\scale]
    \node[anchor=west] at (-9,4.3) {(c) \textbf{ND}+SD};
    \node[circle,fill,inner sep=\PS] at (-1.5,0) (A) {};
    \node[circle,fill=\bcolb,inner sep=\PS] at ( 0,0) (B) {};
    \node[circle,fill=\bcola,inner sep=\PS] at ( 1.5,0) (C) {};
    \node[circle,fill,inner sep=\PS] at ( 3,0) {};
    \node[circle,fill,inner sep=\PS] at (-3,0) {};
    \node[draw, inner sep=5pt] at (+\FPDx, \FPDy) (R) {FPD};
    \node[draw, inner sep=5pt, inner xsep=15pt] at (0, +\CDt) (T) {central detector};
    \node[draw, inner sep=5pt, inner xsep=15pt] at (0, -\CDb) (T) {central detector};
    \draw[->,>=stealth, \bcolb, line width=3pt] (B) -- +(70: 1.6);
    \draw[->,>=stealth, \bcolb, line width=3pt] (B) -- +(-80: 1.6);
    \draw[\bcola, line width=1pt] (R.-170) -- (C);
  \end{tikzpicture}

  \caption{Jet events with one forward proton: (a) single diffractive jets, (b) non-diffractive jets + single diffraction.}
  \label{fig:SDevents}
\end{figure}

As mentioned in the previous section, a measurement of the arrival time of a single forward proton is not useful for discrimination between the signal and background processes.
This is related to the fact that in addition to the position spread of the interactions one deals also with their time spread, which is of the same size (in units of $c=1$).

At the LHC, both ATLAS and CMS experiments foresee upgrades of their central detectors that would provide timing measurement \cite{ATLAStime,CMStime}.
Additional information about the time of the $pp$ interaction will allow the pile-up background rejection also for processes with a measured single forward proton.
Here, only the signal and one background cases have to be considered:
\begin{itemize}
  \item signal -- single diffractive production of jets (\textbf{D}), see Fig.~\ref{fig:SDevents}(a),
  \item background -- non-diffractive production of jets plus one proton from another pile-up interaction (\textbf{ND}+SD), see Fig. \ref{fig:SDevents}(b).
\end{itemize}

Having the measurement of the interaction time, $t_0$, and the interaction vertex position, $z_0$, it is possible to calculate the expected arrival time of the forward proton and compare it to the measured value.
For the signal events, the difference between these times will be equal to:
\begin{linenomath*}\[
  \Delta t_\text{\textbf{SD}} = \delta_t + \delta_z - \delta_{p}
  \]\end{linenomath*}
while for background:
\begin{linenomath*}\[
  \Delta t_\text{\textbf{ND}+SD} = 
  \delta_t + \delta_z - \delta_{p} + t_{0} - t_{1} + z_{0} - z_{1},
  \]\end{linenomath*}
where $\delta_z$ is the error on the interaction position measurement, 
$\delta_t$ and $\delta_p$ are the errors on the central and forward times,
while $z_0$, $t_0$ and $z_1$ and $t_1$ are the positions and 
times of interactions producing jets and the proton, respectively.

\begin{figure}[htbp]
  \centering
  \includegraphics[width=\linewidth,page=2]{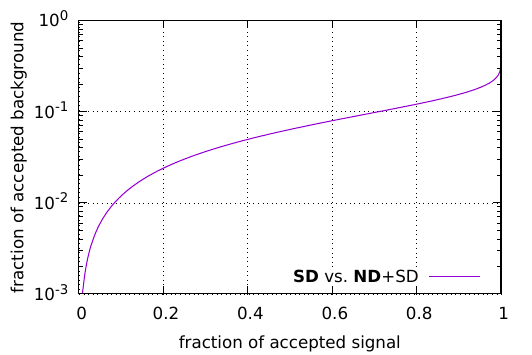}
  \caption{Performance of the signal vs. background discrimination using $\Delta t$ variable.}
  \label{fig:ROCdt}
\end{figure}

The variances of the corresponding $\Delta t$ distributions are:
\begin{linenomath*}\[
  \sigma^2(\Delta t_\text{SD}) = \sigma^2_t + \sigma^2_z + \sigma^2_p
  \]\end{linenomath*}
and 
\begin{linenomath*}\[
  \sigma^2(\Delta t_\text{\textbf{ND}+SD}) = \sigma^2_t + \sigma^2_z + \sigma^2_p + 2\sigma^2_b,
  \]\end{linenomath*}
where $\sigma^2_t$ denote the resolution of the central timing measurement.
For the signal process the variance is a function of the measurement resolutions.
For the background an additional term related to the bunch width appears.
The performance of the discrimination assuming 30 ps resolution of both the central and forward timing systems and 1 mm resolution of the $z$ measurement is presented in Figure \ref{fig:ROCdt}.
The results turns out to be very similar to the SD rejection shown in Figure \ref{fig:ROCdz}.

\newcommand{\Z}{z^\ast}
\newcommand{\T}{t^\ast}

\section{Using central timing for measurements with two protons}

In this section, the combined case is considered where the timing measurement is performed on both protons and on the central state.
It is not straightforward to accommodate a new measurement into the background rejection methods based on calculating $\Delta z$ or $\Delta t$.
Therefore, an alternative approach, which treats all experimental measurements on the same footing, will be applied.
One can define a $\chi^2$ function testing the hypothesis that all measurements correspond to the same interaction taking place at $(\T, \Z)$:
\begin{multline*}
  \chi^2(\T, \Z)  
  =
  \frac{(t - \T)^2}{\sigma^2_t}
  +
  \frac{(z - \Z)^2}{\sigma^2_z} \\
  +
  \frac{(t_+ - L_+ - \T + \Z)^2}{\sigma^2_p}
  +
  \frac{(t_- - L_- - \T - \Z)^2}{\sigma^2_p}
\end{multline*}
The value of this function at the minimum, $\chi^2_\text{min}$, can be used as the discriminating variable.

It is interesting to check what is the outcome of this method when applied to the situations discussed above.
In these cases, not all terms of the $\chi^2$ function are appropriate:
for the double-tag processes, discussed in Section 2, the central timing measurements are not performed, 
while for the single-tag processes, discussed in Section~3, only one proton is measured.
After defining the $\chi^2$ function appropriately, the minimisation can be performed analytically.
For the double-tag measurements, for all four considered processes, the minimal $\chi^2$ value turns out to be
\begin{linenomath*}\[
  \chi^2_\text{min} = 
  \frac{2(\Delta z)^2}{\sigma_p^2 + 2 \sigma_z^2} ,
  \]\end{linenomath*}
where $\Delta z$ is the discriminating variable used in the previous method.
For the single-tag measurement, one finds:
\begin{linenomath*}\[
  \chi^2_\text{min} = 
  \frac{(\Delta t)^2}{\sigma_p^2 + \sigma_t^2+ 2 \sigma_z^2} ,
  \]\end{linenomath*}
This shows that in these cases, the methods presented before are fully equivalent to the $\chi^2$ method.

With the inclusion of the central timing term to the $\chi^2$ function, the analytic calculations of rejection power become rather lengthy, leading to formulae which are difficult to present in a simple form and interpret in terms of well known distributions. 
However, it is possible to perform a numerical Monte Carlo simulation of the problem and to study the gain introduced by an additional measurement.
In the following the three cases will be compared:
\begin{itemize}
  \item[(A)] no central timing (as in Section 2),
  \item[(B)] resolution of central timing equal to resolution of forward timing,
  \item[(C)] infinitely small resolution of central timing.
\end{itemize}
For all the above outlined cases the forward timing resolution of 30~ps will be used.

\begin{figure}[htbp]
  \centering
  \includegraphics[width=\linewidth,page=6]{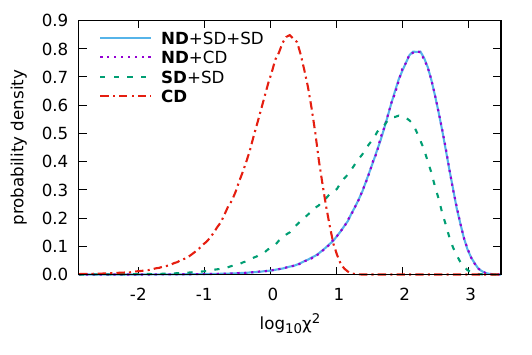}
  \caption{Distributions of $\log_{10}\chi^2_\text{min}$ for the  signal and background processes assuming a 30~ps resolution of 
  the central and forward timing measurements.}
  \label{fig:chi2min}
\end{figure}

Figure \ref{fig:chi2min} presents the distributions of $\log_{10}{\chi^2_\text{min}}$ for scenario (B) for all discussed processes.
A clear discriminating power is observed -- for the signal process the $\chi^2_\text{min}$ values are on average significantly smaller 
than those obtained for the background processes.
However, some overlap of the distribution is present and makes it impossible to reject all the background.
One can also see that the separation is better for the \textbf{ND}+SD+SD and \textbf{ND}+CD background than for the \textbf{SD}+SD, which is true for all 
scenarios.

\begin{figure}[t]
  (a)\\
  \includegraphics[width=\linewidth,page=3]{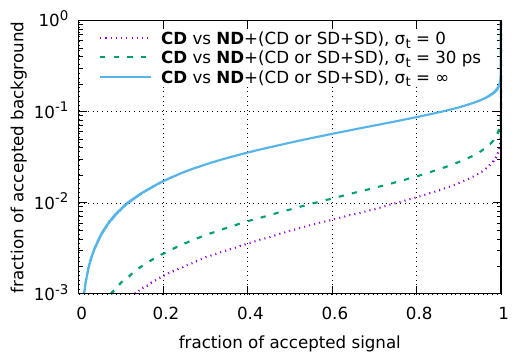}\\
  (b)\\
  \includegraphics[width=\linewidth,page=4]{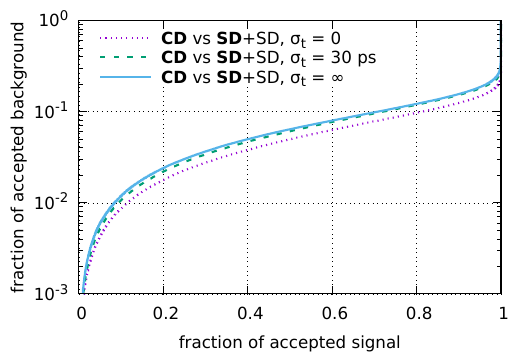}
  \caption{Performance of the signal against background discrimination for three considered scenarios (see text) for (a) \textbf{ND}+SD+SD and \textbf{ND}+CD backgrounds, (b) for \textbf{SD}+SD background.}
  \label{fig:ROCchi2}
\end{figure}

The three considered scenarios are compared in Figure \ref{fig:ROCchi2}, which presents the dependence of the fraction of the 
accepted signal on the fraction of accepted background.
For the \textbf{ND}+SD+SD and \textbf{ND}+CD backgrounds, the improvement by the central timing can reach an order of magnitude.
It is interesting to see that the gain between scenarios (A) and (B) is significantly greater than that between scenarios (B) and (C).
On the other hand, the performance of \textbf{SD}+SD background rejection is much less affected by the presence of central timing.
Here, the gain is about a factor of two for scenario~(C).

\section{Summary and conclusions}

Time-of-flight measurements are an important experimental technique in high energy physics.
This paper discussed their use for measurements of rare diffractive and electromagnetic processes with forward intact protons in proton--proton interactions in high pile-up environment, \textit{e.g.} at the LHC or a future collider.

Three different types of measurements were discussed:
\begin{itemize}
  \item measurement of CD-like processes, with two forward protons and timing measurement on both of them,
  \item measurement of SD-like processes, with one forward proton whose ToF is measured and an additional timing measurement on the central state,
  \item measurement of CD-like processes, with two forward protons and timing measurement on both of them and on the central state.
\end{itemize}
For the first two cases, analytical formulae describing the background rejection power were derived.
These results will be useful for designing future experiments involving similar techniques, since they will allow an easy estimation of the performance of the timing detectors without the need of numerical simulations.

For the last considered case, involving three different timing measurements, obtaining an analytic result was not successful.
Instead, the gain in background rejection due to the additional central timing measurement has been estimated using simulations.
The effect on the main background -- non-diffractive processes -- turned out to be very encouraging and reaching an order of magnitude improvement for the LHC-inspired assumptions.

\section*{Acknowledgements}
This work was supported in part by Polish National Science Center grant no. 2015/19/B/ST2/00989.

\printbibliography

\end{document}